\documentclass[iicol,sn-basic]{sn-jnl}
\usepackage{graphicx}
\usepackage{epsfig}
\usepackage{multirow}
\usepackage{amsmath,amssymb,amsfonts}
\usepackage{amsthm}
\usepackage{mathrsfs}
\usepackage{textcomp}
\usepackage{manyfoot}
\usepackage{booktabs}
\begin{document}
\title[Article Title]{Transient rheology and morphology in sheared nanolayer polymer films}

\author{\fnm{Anna} \sur{Dmochowska}}
\author*{\fnm{Jorge} \sur{Peixinho}}
\author{\fnm{Cyrille} \sur{Sollogoub}}
\author*{\fnm{Guillaume} \sur{Miquelard-Garnier}}
\affil{\orgdiv{Laboratoire PIMM}, 
\orgname{CNRS, Arts et Métiers Institute of Technology, Cnam}, 
\orgaddress{\street{151 Boulevard de l'Hôpital}, 
\city{Paris} \postcode{75013}, \country{France}}}

\abstract{The rheology of coextruded layered films of polystyrene/poly(methyl methacrylate) (PS/PMMA) has been studied with small and large amplitude oscillations at a temperature above their glass transition. While the complex viscosity remains constant over the experimental time window for the micron-sized layered films, a decrease has been observed for the nanolayered films. The rheological behavior has then been correlated to the morphological evolution of the multilayer films: while the nanolayers dewet. Layer breakup followed by retraction and coalescence leading to a lamellar-like blend morphology succeeded by a nodular-like morphology has been evidenced in the nanolayer films, for all compositions and conditions tested. The analysis of the microscopic images of the film cross-sections also provided the droplet size distribution. The nodular morphology is achieved more rapidly when the initial layers are the thinnest at low strains, while at high strains the formation of these droplets is prevented.}

\keywords{Polymer melts, Multilayer films, Rheology, Morphology}
\maketitle

{\centering \includegraphics[width=1.0\columnwidth,keepaspectratio]{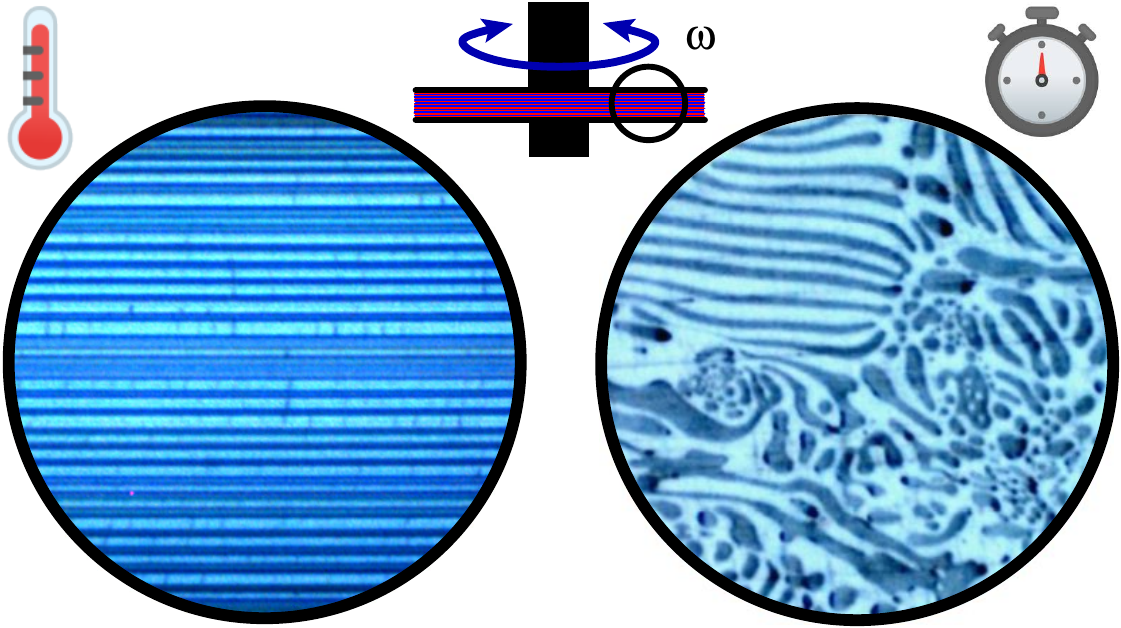}}

\section*{Introduction}

The stability of (ultra-)thin liquid films has been an area of interest for a long time \citep{Vrij1966}, both from a fundamental and applied point of view \citep{deGennes2004}. 
When deposited on a substrate with no affinity, a liquid film will dewet, assuming its thickness is much smaller than the capillary length.
The case of polymer films is of special interest since polymers are complex fluids \citep{Krausch1997} and their use in the form of thin films find many applications \citep{herminghaus2006}, such as coatings, lubricants, etc. 
The constitutive macromolecules, having dimensions $\sim$10 nm, can also give rise to confinement and interfacial effects that become visible at relatively large scale (thickness $\sim$100 nm) \citep{beena2017}. 

The dewetting mechanism can be either spinodal, or through nucleation and growth, but in either case is related to thermal fluctuations of the interfaces \citep{deGennes2004}. 
Both numerical and experimental studies have been performed on the dewetting dynamics, i.e. the speed at which holes grow in the dewetting film after their appearance. 
Several regimes have been identified, depending on the interactions with the substrate, viscosity ratio, film thickness \citep{Krausch1997}.
However, fewer experimental studies deal with the characterization of the dewetting onset due to measurement difficulties, such as low resolution of an optical microscope, which does not allow capturing the hole's nucleation, or too low scanning speed of an atomic force microscope (AFM) to perform in-situ measurements \citep{Reiter2005}.\\
An interesting case study concerns multi-nanolayer films. 
Contrary to the previously discussed monolayer films, the total thickness in these systems is on the order of 10-100 $\mu$m, but made of thousands of alternating layers of two (usually immiscible) polymers superposed in a one-step process, multilayer coextrusion. 
Each constitutive layer has a nanometric thickness. 
These systems can find applications in several domains \citep{Baer2017, Langhe2016}, such as packaging, as it has been for example shown that confinement can lead to in-plane crystallization more favorable to gas barrier properties \citep{wang2009}. 
Because of the high temperatures involved during the extrusion process, the layer stability in these systems has been studied extensively as a function of processing parameters \citep{Bironeau2017,Feng2018}, and model experiments derived from those on ultra-thin (monolayer) films have been proposed to study the dewetting dynamics taking into account the polymer-polymer interfaces \citep{Zhu2016, Chebil2018}.\\
However, the role of shear in layer or thin-film stability has not been fully elucidated yet \citep{Craster2009}. 
Shear commonly occurs during industrial processes for fabricating thin films or layers: multilayer coextrusion, flow-coating, etc. 
Shear has been anticipated, rather counter-intuitively, to delay or even suppress rupture in free-standing polymer films \citep{Kadri2021,Davis2010}. 
These theoretical findings, based on 2D simulations, have been contradicted by a recent 3D study \citep{Dhaliwal2024} in which the global stability of the system is only slightly improved, even at high shear levels, because even though shear will lower interfacial perturbation along the direction it is applied, hole formation can still occur perpendicular to shear. 
Nonetheless, this hypothesis of stabilized nanolayers during processing due to shear has been proposed in the case of multilayer coextrusion in several studies  \citep{Dmochowska2022,Bironeau2017} but until now, no experiment of our knowledge has been able to definitely prove it.

Since Palierne's seminal work \citep{Palierne1990,Grmela2001}, rheological measurements have been used to study various interfacial phenomena in immiscible polymer blends, and more recently on multi-nanolayer films, such as interfacial compatibilization \citep{Beuguel2019, Beuguel2020}, interfacial slip \citep{Zhao2002, Li2024} or interfacial elasticity \citep{Jordan2019, Dmochowska2023}. 
These studies have shown that multi-nanolayer films, due to the well-controlled multiplication of interfaces, can be used as model samples to enhance such interfacial phenomena as well as confinement effects in coextruded polymer materials \citep{Liu2005}. More generally, rheology is known to be a sensitive probe of the morphology in immiscible polymer blends \citep{Lee1994,Martin2000}.
Many studies have focused on the morphology/rheology interrelationships in immiscible blends such as PS/PMMA \citep{Calvao2005,Weis1998}, while others used rheology as a probe of morphology changes in blends especially in the case of reactive compatibilization \citep{Huo2007,Yee2007,Afsari2020}.
Rheological characterization is thus expected to be a relevant tool to track morphological changes in multi-nanolayered films induced by layer breakups.
In consequence, we propose here the use of coupled rheological and microscopy, both optical and AFM, of initially stable multi-nanolayer films, as a non-direct method for capturing the layers' dewetting as a function of temperature and shear, in relation with realistic processing conditions.

\section*{Materials and Methods}

\subsection*{Materials}

Following earlier works \citep{Dmochowska2022,Dmochowska2023}, polystyrene PS 1340 from Total and poly(methyl methacrylate) PMMA VM100 from Arkema were selected to produce multilayer films. 
Their glass transition temperatures are 96 and 94 \textdegree C, respectively. 
The molecular weights and densities have been determined previously \citep{Bironeau2017,Zhu2016}. 
These two polymers were chosen because their viscosity ratio is close to unity at the processing temperature, which facilitates multilayer coextrusion without macroscopic instabilities \citep{Dooley2002}. 
The viscoelastic properties of the neat materials have been characterized thoroughly elsewhere \citep{Dmochowska2023}.

\subsection*{Film fabrication}

\begin{figure*}
\centering
\includegraphics[width=1\linewidth]{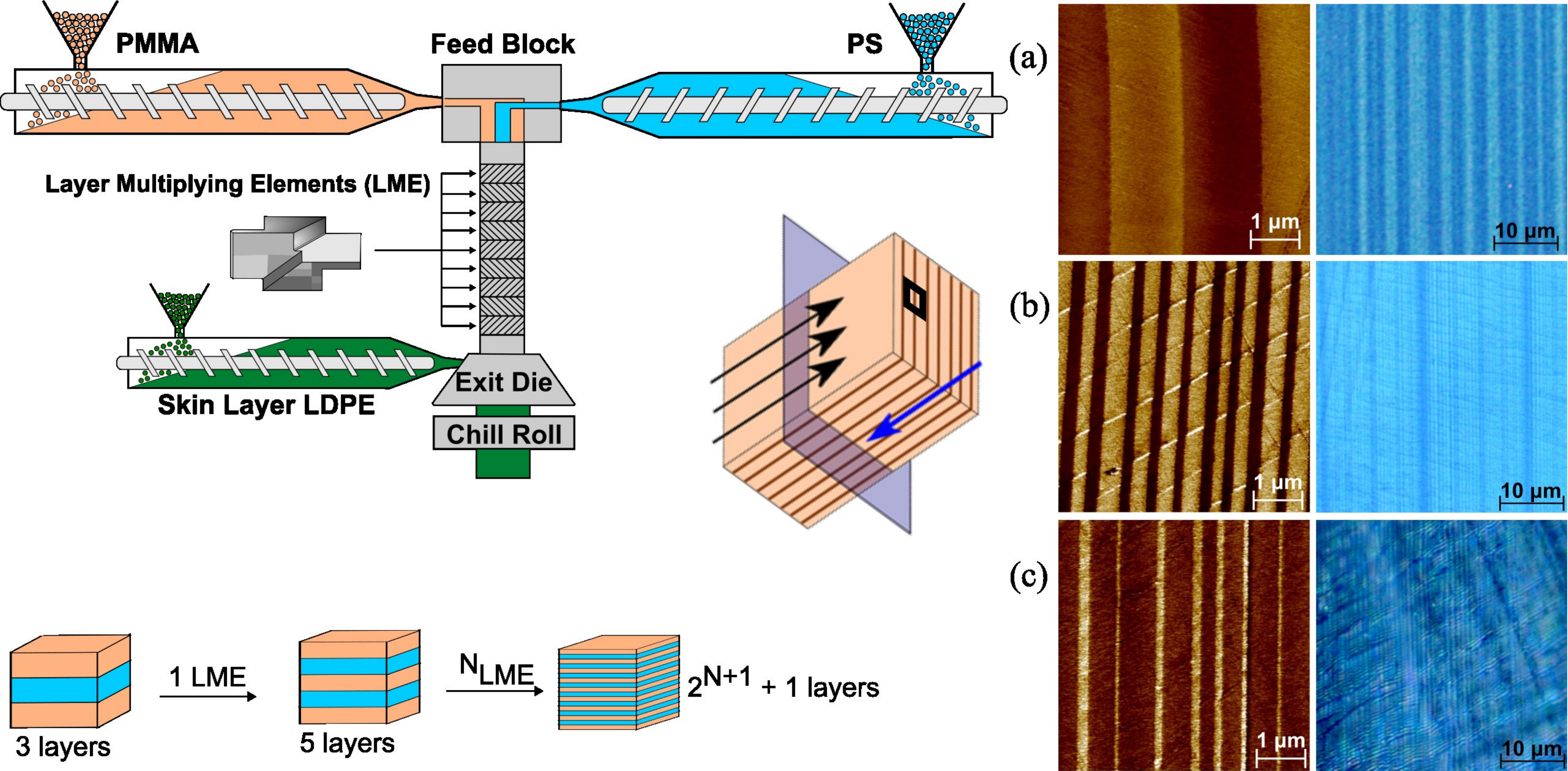}
\caption{Scheme of the multi-nanolayer coextrusion, the layer multiplying element operation and microscopy observation. 
Cross-sectional AFM (left column) and optical microscopy images (right column) of the films: (a) 129 layers 60/40 PS/PMMA; (b) 2049 layers 60/40 and (c) 2049 layers 30/70.} 
\label{fig1}
\end{figure*}

PS/PMMA multilayer films were fabricated using a lab-made customized multilayer coextrusion line \citep{Montana2018}, schematized in Fig. \ref{fig1}. 
Again, a more detailed description of the processing of the films can be found in \citep{Dmochowska2023} but will be described briefly in the following. 
The melts of two polymers enter a feed block with a controlled throughput of about 4 kg/h. 
The polymer flow is split vertically, spread horizontally, and then recombined at each layer multiplying element (LME) encountered. 
At the exit die, a sacrificial layer of melted low-density polyethylene, LDPE 1022 FN from Total, is added to improve the surface homogeneity and reduce the thickness of the multilayer film. The film is collected at a 90 \textdegree C chill roll, with the lowest possible draw ratio $\sim$ 1.3 to
minimize as much as possible post-extrusion chain relaxation during the rheological experiments. 
The skin layer of LDPE has no adhesion to the multilayer films and is removed before any further experiments. 
The films used in this study were produced with 5 and 10 LMEs, leading to films having 129 and 2049 layers respectively, with a composition of 60/40 wt\% PS/PMMA. An additional composition of 30/70 wt\% was studied for the film with 2049 layers.

\subsection*{Film morphology}

The morphology and individual layer thicknesses of the fabricated multilayer films were characterized with a transmission optical microscope (OM) Axio Imager 2 (Zeiss) and an AFM Nanoscope V (Veeco), to reveal the layered morphologies of the films. 
The individual layer thickness, for example the thickness of PS, $e_{\text{\tiny{PS}}}$, can be estimated as follows:
\begin{equation}
e_{\text{\tiny{PS}}} = H_{\text{\tiny{M}}} \times \frac{\varphi_{\text{\tiny{PS}}}}{n_{\text{\tiny{PS}}}}
\label{eq1}
\end{equation}
with $H_{\text{\tiny{M}}}$ the total thickness of the multilayer film, $\varphi_{\text{\tiny{PS}}}$ the volume fraction of PS in the film and $n_{\text{\tiny{PS}}}$ the number of PS layers. A similar equation can be written for PMMA. 

The samples were taken from the middle of the extruded film parallel to the extrusion flow, then cut first with scissors to reveal their cross-section before slicing using an ultramicrotome. 
The first slices were made with a glass knife, followed by slices with a diamond knife (Diatome Ultra 45\textdegree) to achieve the flattest surface possible. 
Figure \ref{fig1}(a-c) shows the morphology of the chosen samples ``as-extruded", observed at different scales by AFM and OM. 
The layer thicknesses of films with 129 layers were measured following the method developed in \citep{Bironeau2016}. 
Their experimental layer thicknesses are $e_{\text{\tiny{PS}}}=1.8\pm0.7$ $\mu$m and $e_{\text{\tiny{PMMA}}}=1.4\pm0.6$ $\mu$m, in good agreement with the calculated layer thicknesses from \eqref{eq1}: 1.6 and 1.4 $\mu$m for PS and PMMA, respectively \citep{Dmochowska2023}. 

The film with 2049 layers and composition 60/40 PS/PMMA had layer thicknesses of $e_{\text{\tiny{PS}}}=287\pm91$ nm and $e_{\text{\tiny{PMMA}}}=221\pm78$ nm, again in good agreement with the values obtained from \eqref{eq1} (298 and 260 nm). 
The film with composition 30/70 PS/PMMA had layer thicknesses of $e_{\text{\tiny{PS}}}=132\pm60$ nm and $e_{\text{\tiny{PMMA}}}=391\pm91$ nm (calculated values were 106 and 376 nm, respectively) \citep{Dmochowska2023}. 

\subsection*{Rheology}

Oscillatory shear experiments were performed using a rheometer (DHR 20, TA Instruments) with 25 mm plate-plate geometry at 180 \textdegree C under air flow to mimic real processing conditions. 
The multilayer films were cut into disks with a diameter of 25 mm. 
To facilitate the cutting, the films were stored beforehand in an oven at 100 \textdegree C, a temperature close to the glass transition temperature of PS and PMMA for about 1 minute. 
Note that in the 129 layers case, 15 films were stacked on top of each other to compare samples with a similar total number of layers ($15\times129=1935$ versus 2049).
The gap between the plates in the rheometer was set according to the total thickness, $H_M$, of each multilayer film, which was verified by a digital length gauge system with an accuracy of 0.2 $\mu$m (Heidenhain).
The temperature, 180 \textdegree C, was chosen according to previous results for the dewetting speed of a PS layer embedded between two thicker PMMA layers \citep{Dmochowska2022}, in order to define a reasonable experimental time ($\sim 1$ h). 
Here, oscillatory shear experiments were preferred to steady shear ones due to slip  at high shear in the latter case.
There are two parameters to consider when performing oscillatory shear: the oscillatory pulsation, $\omega$, and the shear strain, $\gamma$. 
For small amplitude oscillatory shear (SAOS), the shear strain was set to $\gamma=0.1$\% while $\omega=0.1$, 1, 10 or 100 rad/s. 
The measurements were monitored as a function of time.
At $t=3$, 10 and 30 min, samples were quickly cooled down or temperature quenched using an air gun and delicately taken out of the rheometer oven to avoid further changes in their morphology before microscopy observation. 
Large amplitude oscillatory shear (LAOS) measurements were performed with $\gamma=10$\% at $\omega=1$ and 100 rad/s, and $\gamma=100$\% at $\omega=1$ rad/s for one chosen multilayer film having 2049 layers and composition 60/40 PS/PMMA. 

In addition, measurements in ``static'' mode were performed for comparison, where a film with dimensions $10\times10$ mm was cut out of the multilayer film and placed between microscope glass cover slides.
These samples were then inserted into a hotstage (Linkam, THMS600), also heated up at 180 \textdegree C for $t=3$, 10 and 30 min.

\section*{Results and discussion}

\subsection*{Effect of $\omega$ on SAOS}

\begin{figure*}
\centering
\includegraphics[width=1.0\linewidth]{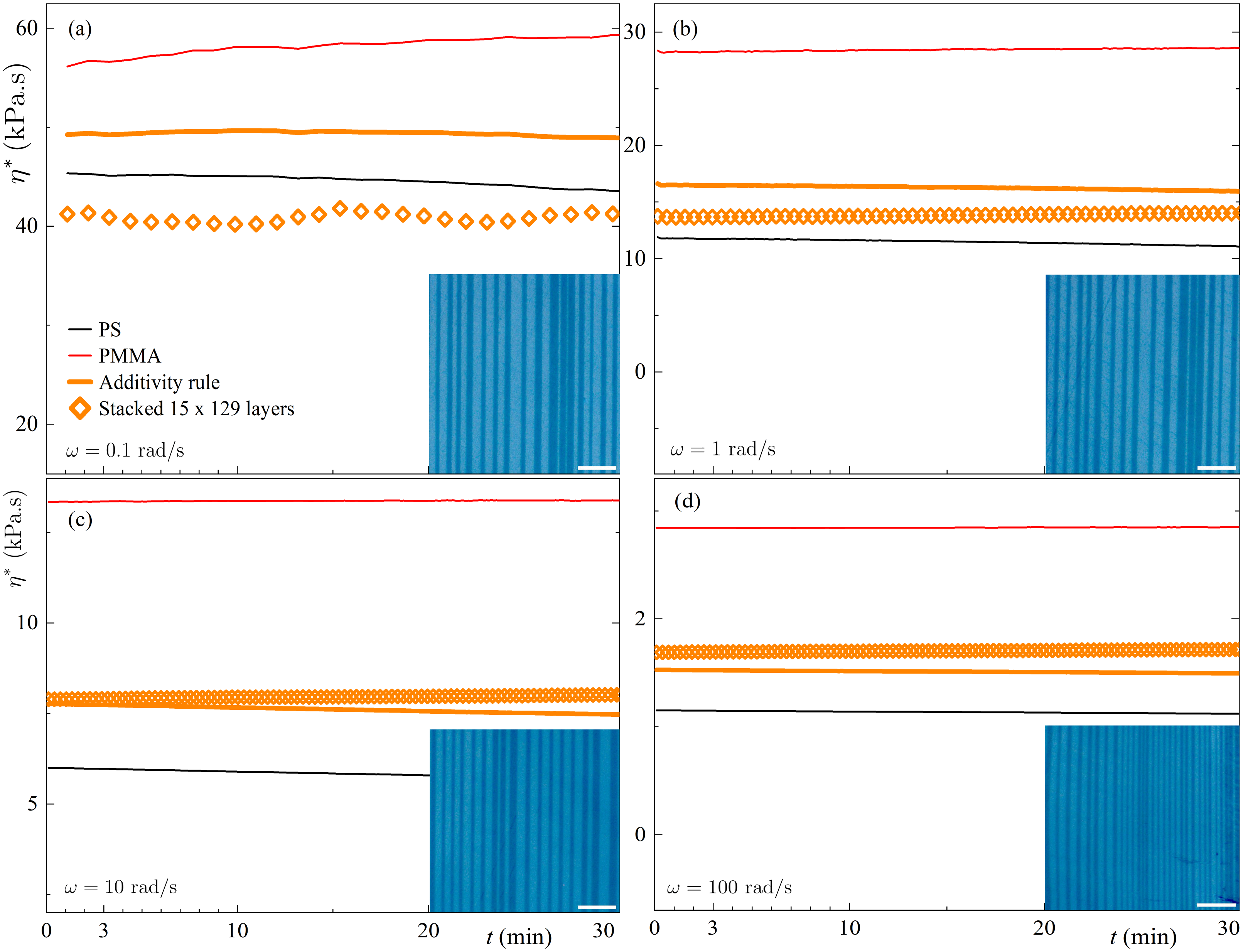}
\caption{The complex viscosity, $\eta^*$, in linear scale as a function of time at 180 \textdegree C for PS, PMMA and the stacked multilayer film ($15\times129$ layers, so a total of 1935 layers) at shear strain $\gamma=0.1$\%  for (a) $\omega=0.1$, (b) 1, (c) 10 and (d) 100 rad/s. The orange symbols are the measured data while the orange thick line is the viscosity calculated from the additivity rule. The inset images show cross-sections optical microscopic observations of the preserved layers at $t=30$ min with the scale bar representing 10 $\mu$m. Note the samples' cross-sections are presented vertically for easy comparison with the scale bar, yet the layers were parallel to the rheometer plates.} 
\label{fig2}
\end{figure*}

The effect of $\omega$ on layer stability was first tested in multilayer films with 129 layers (thicker layers). 
In Fig. \ref{fig2}, it is clearly seen that the complex shear viscosity, $\eta^*$, is constant over time for all tested $\omega$.  
At 180 \textdegree C, the viscosity is shear thinning as shown previously, with shear thinning index of $n=0.25$ for PS and $n=0.35$ for PMMA \citep{Dmochowska2023}.
Assuming the Cox-Merz principle is valid for nanolayered blends as it is classically done for other immiscible blend morphologies \citep{Charfeddine2020}, this explains why the viscosity values decrease with $\omega$.

The experimental viscosity for the multilayer film can be compared to the one estimated from a ``simple'' additivity rule, which takes into consideration the measured viscosities of PS and PMMA and their respective volume fractions, $\varphi$, in the multilayer film:
\begin{equation}
\frac{1}{\eta^*}=\frac{\varphi_{\text{\tiny{PS}}}}{\eta_\text{\tiny{PS}}^*}+\frac{\varphi_{\text{\tiny{PMMA}}}}{\eta_\text{\tiny{PMMA}}^*}.
\label{eq2}
\end{equation}
We note that this is the simplest additivity rule model, not taking into account the possible effect of interphases \citep{Lee2009,Jordan2019} which would add an additional contribution to $\eta^*$ in proportion of the volume fraction of interphase. Since this volume fraction of interphase remains small in the range of thicknesses studied (see our detailed study of this effect in \cite{Dmochowska2023}), neglecting it in this specific case will not change the conclusions. 
For all $\omega$, there is a reasonable agreement between the simple additivity rule model and the measured viscosity, which remains on the order of 15\%. 
Both the model and the experiments display a constant viscosity value over time.

After $t=30$ min, the samples' cross-sections were imaged with OM and are shown as insets on each subplot of Fig. \ref{fig2}. 
For these layer thicknesses, no significant morphology changes are observed, as the layers remain continuous.

To explain these results, we can come back to the simple case of a single liquid thin layer of thickness $e$ deposited on an immiscible liquid substrate described by  \cite{BrochardWyart1993}. A critical layer thickness can be defined as $e_i\approx\theta_e\sqrt{\gamma/\rho g}$, where $\theta_e$ is the equilibrium contact angle, $\gamma$ is the interfacial tension between PS and PMMA, $\rho$ the density and $g$ the gravitational acceleration. Using data available from the literature (see for example \cite{Zhu2016}), we obtain $e_{\tiny{c}}\approx 70$ $\mu$m. Below this thickness, the layer is in a metastable state and will dewet through nucleation and growth of holes starting from defects or dusts initially present in the film. When the thickness becomes on the order of $e_i\approx\sqrt{A_{\tiny{H}}/3\pi\gamma}\approx\ 10$ nm \citep{Bironeau2017} where $A_{\tiny{H}}$ is the Hamaker constant, the system becomes unstable and spinodal dewetting is expected. 
Thus, even though these 129 layers films with layer thicknesses $\sim 1\,\mu$m are expected to dewet, they were stable in the experimental time window chosen at this temperature. 
In the spinodal regime, \cite{Vrij1966} has shown that the rupture time scales as $\eta e^5$, meaning that for this specific system in these conditions, thicknesses of 1 $\mu$m yield typical rupture time of several days while layers having 100 nm thicknesses shall rupture within a few tens of seconds. Hence, to observe significant morphology change in the 129 layers films, experiments would require a significantly longer time (probably non-achievable experimentally even though at these thicknesses the dewetting mechanism is nucleation dominated) or a much higher temperature to lower drastically the viscosity.

\begin{figure*}
\centering
\includegraphics[width=1.0\linewidth]{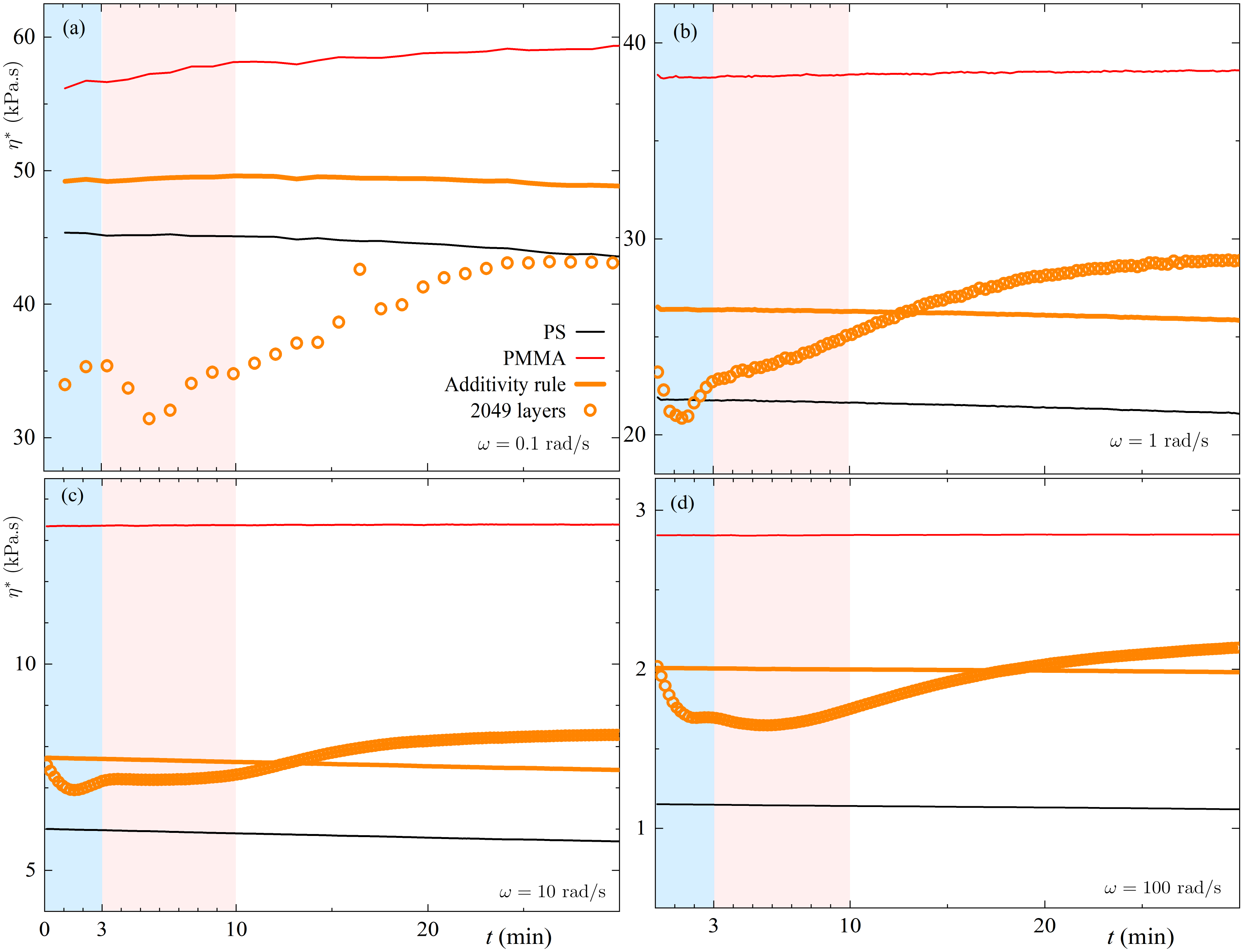}
\caption{The complex viscosity, $\eta^*$, as a function of time at 180 \textdegree C for PS, PMMA and the 2049 layers 60/40 film at shear strain $\gamma=0.1$\% for (a) $\omega=0.1$, (b) 1, (c) 10 and (d) 100 rad/s. The orange symbols are the measured data while the orange thick line is the viscosity calculated from the additivity rule.} 
\label{fig3}
\end{figure*}

The 2049 layers 60/40 films were then tested, as shown in Fig. \ref{fig3} again at $\gamma=0.1\%$ for $\omega=0.1$, 1, 10 and 100 rad/s.
Here, the viscosity is not constant over time, contrary to the previous case. 
More specifically, three time zones may be distinguished, represented by shaded colors. 
At the beginning, there is a sudden decrease of $\eta^*$. 
The minimum value of $\eta^*$ is reached at around $t\approx2$ min for all tested conditions, which in some cases is followed by a short increase that stops at $t\approx3$ min. 
The next time zone is until $t=10$ min where either a short plateau or a small increase of $\eta^*$ is observed. 
Then, $\eta^*$ slightly increases until a second plateau is reached toward the end of each measurement. 
Note that the same behavior was observed for 30/70 films, see Fig. S1 in the Supplementary Information (SI).
Let us now try to correlate these variations with morphological changes.

\begin{figure*}
\centering
\includegraphics[width=1.0\linewidth]{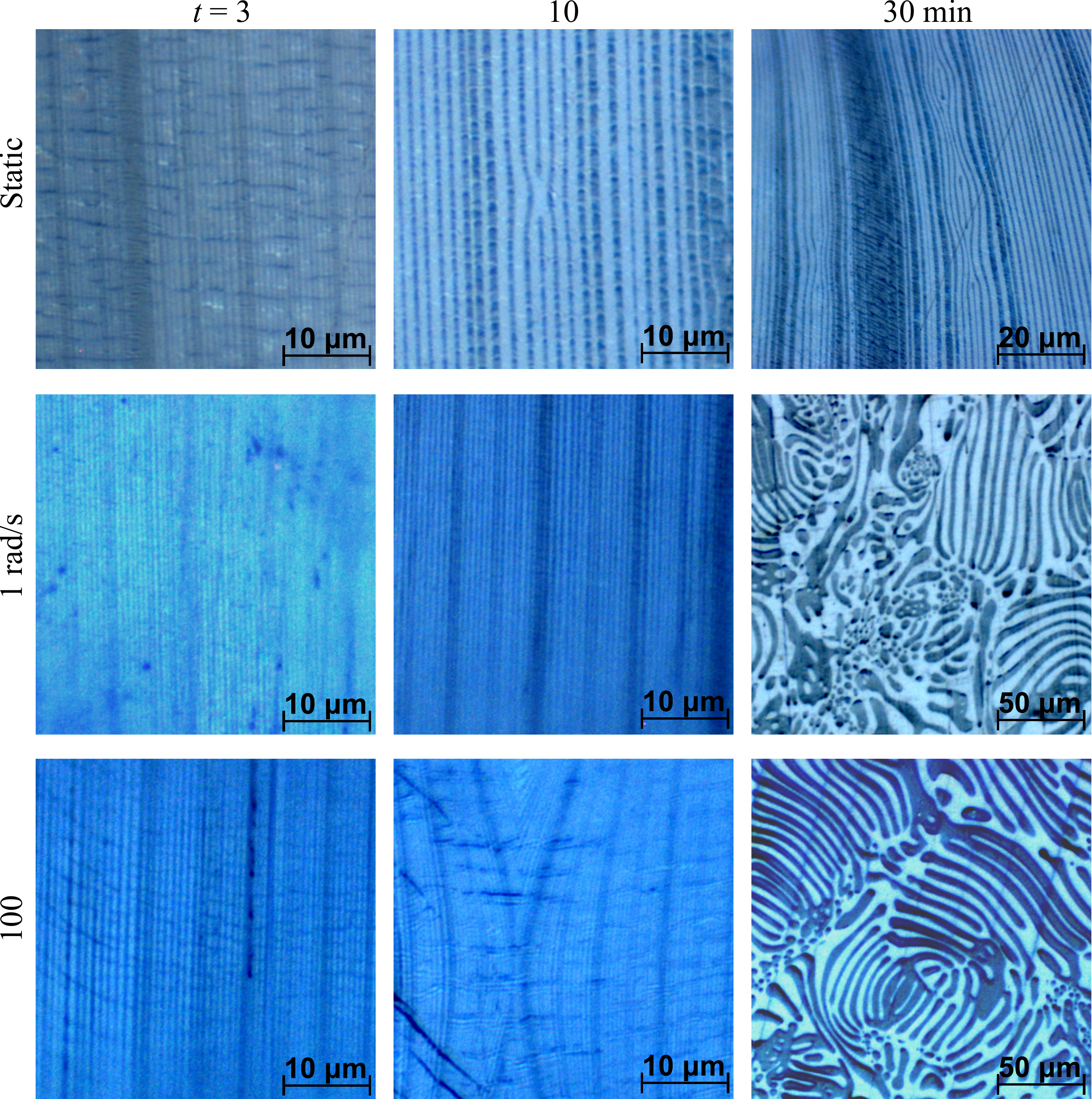}
\caption{OM images of quenched cross-sections of the 2049 layers 60/40 film at different times and pulsations. First line, static conditions. Following lines, cross-sections after oscillating shear at $\gamma=0.1\%$ and $\omega=1$ and 100 rad/s after 3, 10, and 30 min (first, second, and third column, respectively). Light blue represents PS and dark blue PMMA.} 
\label{fig4}
\end{figure*}

OM images of the 2049 layers 60/40 film sheared at different $\omega$ until $t=3$, 10 and 30 min are then shown on Fig. \ref{fig4}. 
For the static case, no changes were observed at $t=3$ min.
After $t=10$ min, the layers became thicker, which may be caused by the sample shrinking due to the polymer chain relaxation, permitted by the fact that the samples are unconstrained in these specific experiments. As expected from Vrij's approximation discussed previously, a small number of PMMA layer breakups can be observed in some locations (see for example in the middle of the image of the static case), which causes the neighboring layers to move closer to one another. 
At $t=30$ min this phenomenon is more pronounced, leading to a waviness of the layers throughout the sample. 
For $\omega=1$ and 100 rad/s, because of the constant gap applied, the layers' coarsening is less pronounced, but some broken or wavy layers can be observed after $t=10$ min.
However, at $t=30$ min, a highly disorganized morphology resembling a co-continuous or lamellar blend morphology \citep{Macosko2000} appears, with dimensions much bigger than the thickness of the initial layers (note the change in the scale bar). Microscopic droplets having various shapes and different levels of circularity are also observed.

If we come back to the comparison with the previous films, it is now clear that though both layer thicknesses shall lead to layer breakup, 129 layers films are stable during the experimental window at all pulsations, while the 2049 layers films experience dewetting with first visible morphology changes at 10 min and a totally altered microstructure after 30 min, especially for the sheared samples. 

Figure S2 in SI shows additional images for $\omega=0.1$ and 10 rad/s, and Fig. S3 in SI shows the same experiments for the film with 2049 layers and 30/70 composition. 
Similar morphology elongated patterns are observed at low $\omega$ for the two compositions. However, for the 30/70 composition at high $\omega$, less elongated areas and more droplets can be seen, which will be further discussed in the following.

\begin{figure*}
\centering
\includegraphics[width=0.70\linewidth]{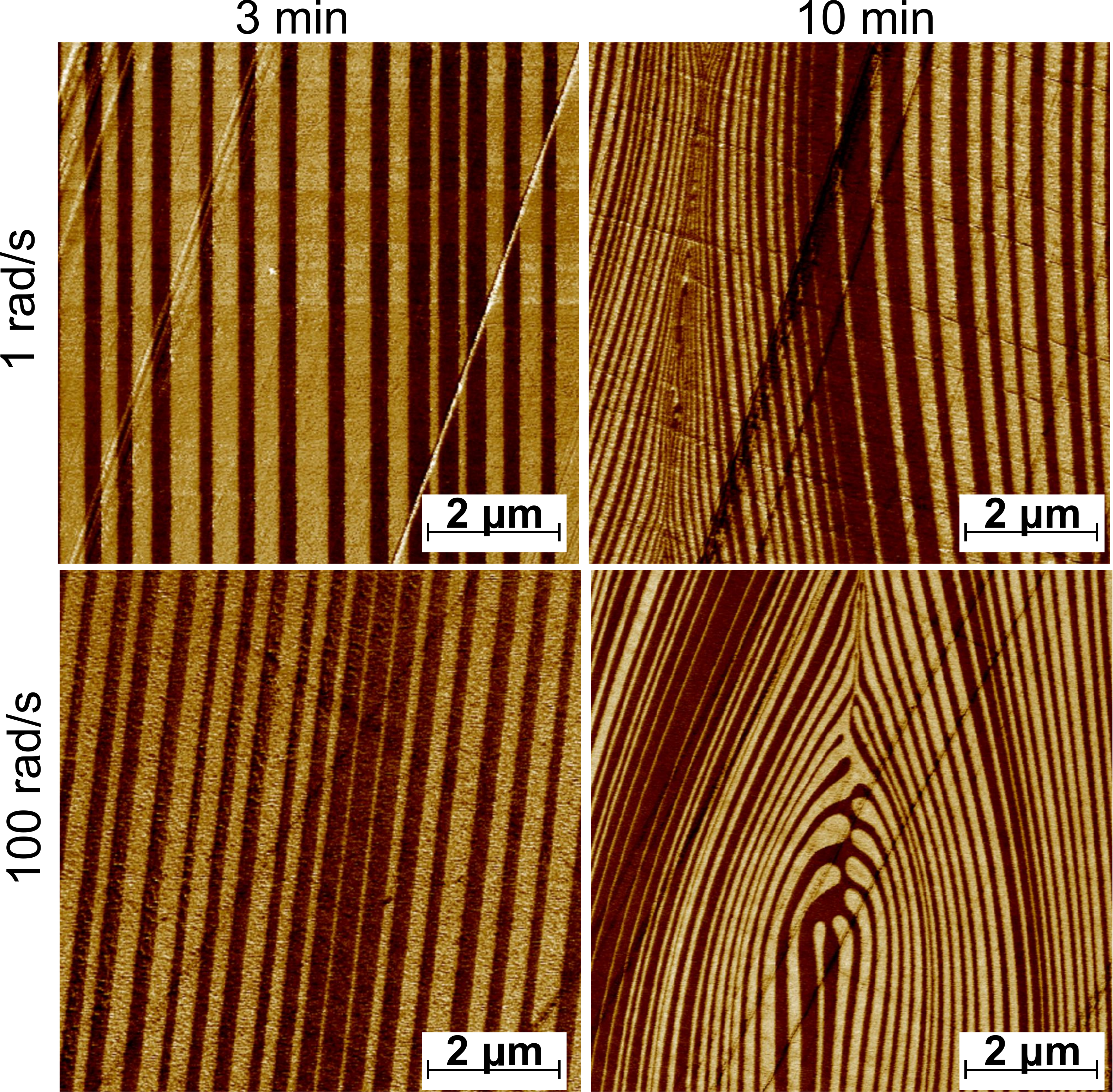}
\caption{AFM images of quenched cross-sections of the 2049 layers 60/40 film at different times and pulsations.} 
\label{fig5}
\end{figure*}

A closer examination of the layers was performed with AFM again in the extrusion direction, as shown in Fig. \ref{fig5}, where yellow and brown colors correspond to PS and PMMA, respectively. Similar observations can be made in the transverse direction \citep{bironeau2016thesis}.
At $t=3$ min, the layers remain unbroken, confirming the OM observations.
For $\omega=1$ and 10 rad/s and $t=10$ min, broken ends layers exhibit elongated and large rims collecting the dewetted polymer with circular cross-section larger than the layer thicknesses \citep{Reyssat2006}.  
This induces flows to rearrange and deform neighboring layers and even leads to layer coalescence in other regions in the image. 
As the holes within the layers grow due to interfacial tension drawing the rim, lamellae get shorter and thicker over time (as seen in Figs. \ref{fig4}, S2 and S3). 
At longer times, rims from different holes in the same layer will coalesce to form cylinder-like structures, which will end up as droplets such as those observed in Fig. \ref{fig4} due to Plateau-Rayleigh type instabilities \citep{Sharma1996}.

\begin{figure*}
\centering
\includegraphics[width=1.0\linewidth]{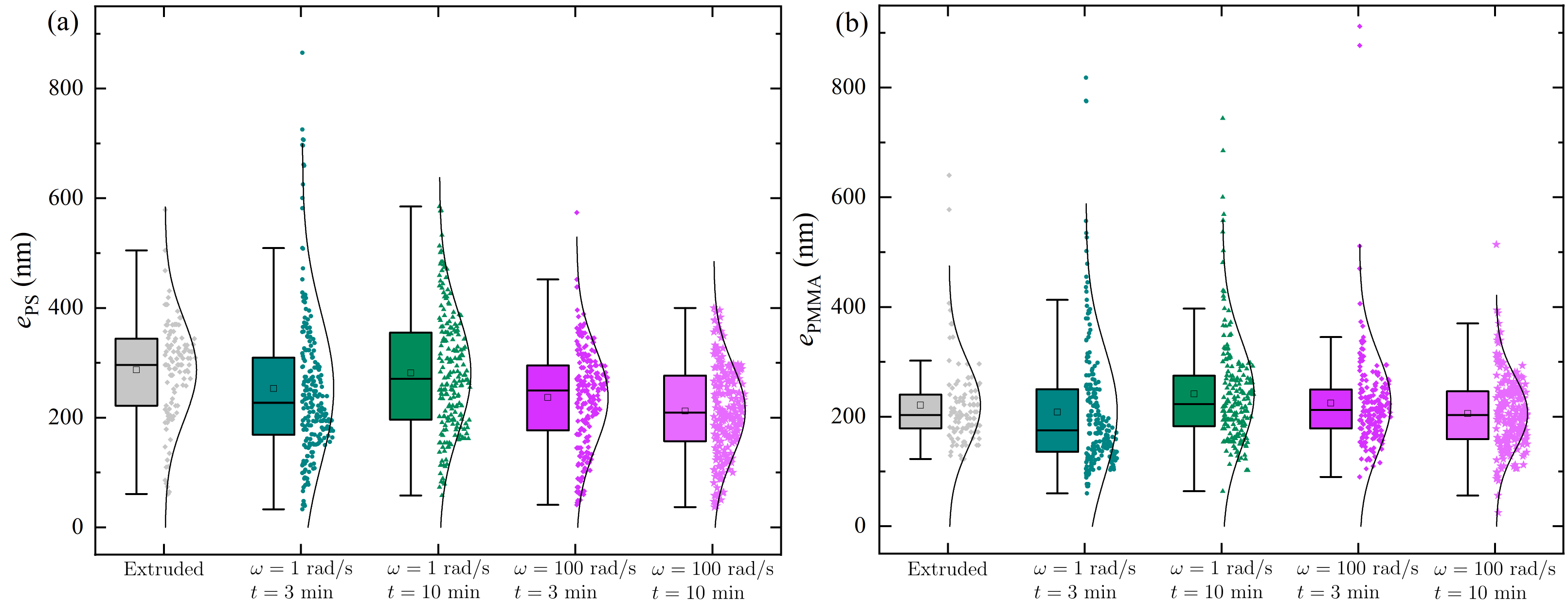}
\caption{Box plot with normal distribution of the layer thicknesses of (a) PS and (b) PMMA measured from AFM images of the 2049 layers 60/40 film, as-extruded films and after shearing at $\omega=1$ and 100 rad/s at $t=3$ and 10 min. The box indicates the upper and lower quartiles of the distribution. The whiskers are determined by the 5$^{\rm th}$ and 95$^{\rm th}$ percentiles. The horizontal line and the open square inside the box represent the median and the mean values, respectively.} 
\label{fig6}
\end{figure*}

Using multiple AFM images, layer thickness measurements were conducted at $t=3$ and 10 min for $\omega=1$ and 100 rad/s.
The layer thickness statistics (based on 100 to 200 layers) are displayed in Fig. \ref{fig6}, where the normal distribution of layer thicknesses remains essentially unchanged compared to the as extruded films, despite few observed breakups (note that the thickness was measured only on layers that were continuous), confirming the observations made in Fig. \ref{fig4}.
Hence, a large number of breakups is needed to significantly modify the mean layer thickness.

As stated previously, under no or low shear a micronic nodular morphology should be the final equilibrium state of the dewetting process, so the droplet size distribution after 30 minutes of shearing was studied next.
To do so, optical microscope images were analyzed using the ImageJ software \citep{Schneider2012} taking into account the image size and resolution, meaning the areas smaller than 5 $\mu{\rm m}^2$ were not considered. A particle tracking analysis was performed on a minimum of 5 images taken at different locations in the sample with magnification $20\times$, therefore covering an area equal to $470\times630$ $\mu$m, about 1/4 of the total area of the sliced sample. Only areas representing more than 1\% of the total PS area are shown in the figures \ref{fig4}, \ref{fig5}, S2 and S3.
First, the area distribution for all shapes of the dispersed phase is presented in Fig. S4, and is similar in shape for all conditions, see also Fig. S5(a) in SI for the 30/70 PS/PMMA. It can be noted that the threshold at which the covered area becomes lower than 1\% occurs around 80 $\mu{\rm m}^2$ for the 60/40 composition, and 50 $\mu{\rm m}^2$ for the 30/70 one, suggesting a sharper distribution of droplets in this latter case. 
In a second approach, assuming large elongated domains are transient and will later become droplets, only areas with circularity (defined as $c=4\pi\mathcal{A}/\mathcal{P}$ with $\mathcal{A}$ the area considered and $\mathcal{P}$ the perimeter of the considered area), $c>0.5$, were considered, i.e. ellipses with radius ratio smaller than 2.
 
\begin{figure}
\centering
\includegraphics[width=1.0\linewidth]{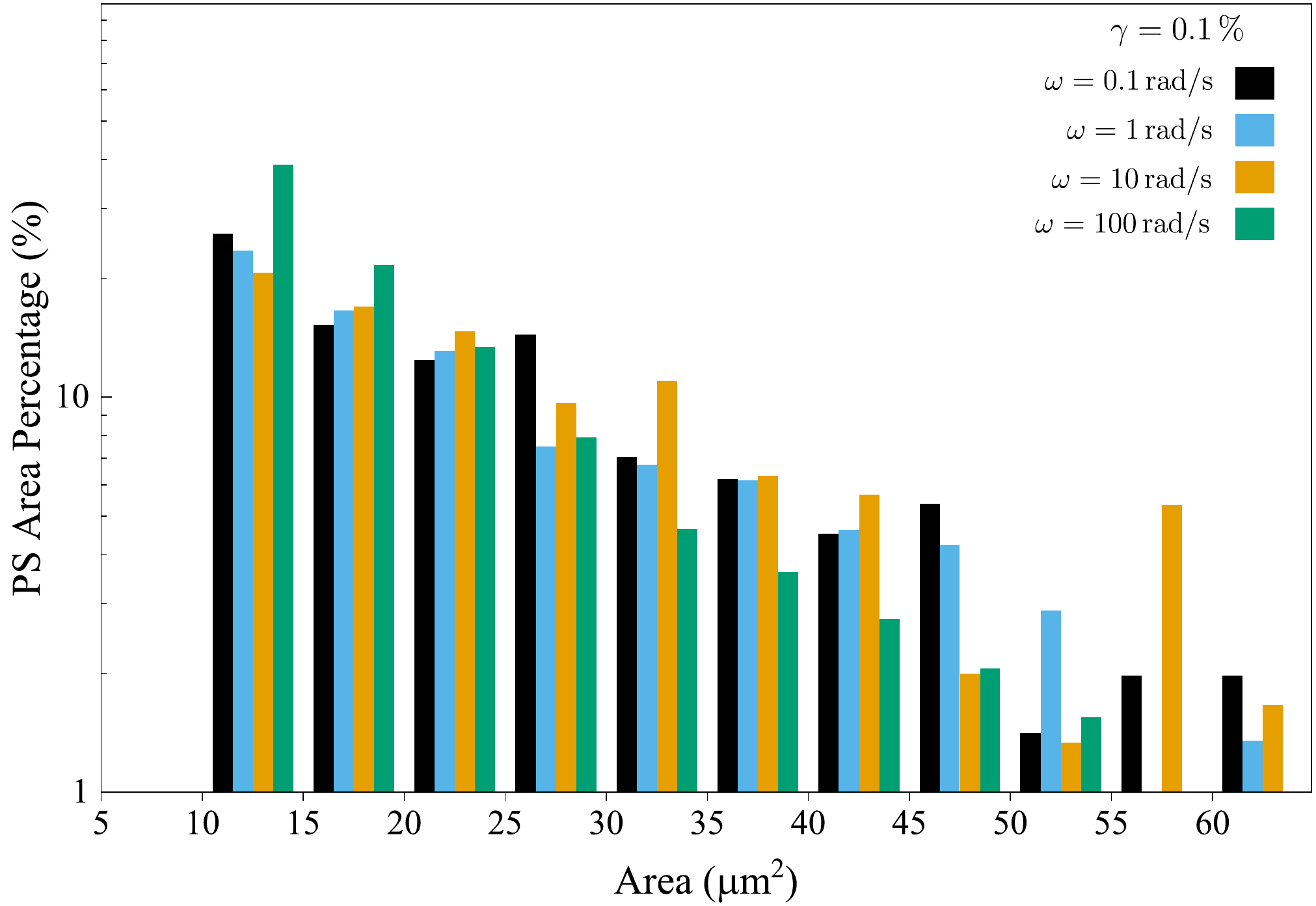}
\caption{Area distribution of PS drops with circularity $c>0.5$ in the cross-section of 2049 layers 60/40 films after $t=30$ min for $\gamma=0.1\%$ and $\omega=0.1$, 1, 10 and 100 rad/s. The bin size of each sample is 5 $\mu$m$^2$.} 
\label{fig7}
\end{figure}

Again the PS area size distribution is similar for all pulsations, as shown in Fig. \ref{fig7}. 
The oscillatory shear is creating droplet size distribution that decreases with increasing area, with a predominance of drops corresponding to an area of 10-15 $\mu$m$^2$, also seen for the 30/70 composition (see Fig. S5(b)) which again displays a narrower distribution than the 60/40 one.
For better comparison, the area percentage of PS drops with circularity $c>0.5$ within the PS dispersed phase (all shapes) is summarized in Tab. \ref{tab1} (first 8 lines) for the 2 compositions and all pulsations. 
The area percentage is below 10\% and does not follow a specific trend for the 60/40 composition with varying  $\omega$, but is significantly higher for the 30/70 composition. Moreover for this composition it clearly increases with $\omega$ (from 16\% to 38\%), confirming what was observed when comparing Fig. \ref{fig4} and Fig. S3. Recalling again that the final stage of dewetting shall be a nodular morphology, this suggests that the dewetting is further advanced at $t=30$ min for the 30/70 composition than for the 60/40 one. This can be explained by the fact that the initial PS layers are thinner in the 30/70 films compared to the 60/40 ones, as well as by the initial difference between PS and PMMA layer thicknesses in the 30/70 films, promoting the breakup of the thinner PS layers first.

\begin{table}
\caption{Summary of the composition, the oscillation parameters and the percentage of drops with circularity $c>0.5$ formed at $t=30$ min}
\centering
\begin{tabular}{lrrrr}
\hline
PS/PMMA &  $\omega$ & $\gamma$ & Percentage area  \\
wt.\%/\% & rad/s & \% & of PS drops \%\\
\hline
60/40 & 0.1 & 0.1 & 6 \\
60/40 & 1 & 0.1 & 8 \\
60/40 & 10 & 0.1 & 6 \\
60/40 & 100 & 0.1 & 9 \\
30/70 & 0.1 & 0.1 & 16 \\
30/70 & 1 & 0.1 & 21 \\
30/70 & 10 & 0.1 & 38 \\
30/70 & 100 & 0.1 & 38 \\
60/40 & 1 & 10 & 6 \\
60/40 & 100 & 10 & 5 \\
60/40 & 1 & 100 & 3 \\
\hline
\end{tabular}
\label{tab1}
\end{table}

Having described the morphology, it is now important to relate it to the rheological findings for both compositions. 
From the observations, the initial drop in $\eta^*$ around $t=3$ min (in Fig. \ref{fig3}) cannot be correlated quantitatively to an onset of dewetting. Still, it is clear that a significant modification of the microstructure occurs during the first 10 min (where the viscosity recovers a value close to its initial one) because of first dewetting events (few holes formation in the layers). 
The end plateau with a viscosity value similar to the additivity rule prediction may also be correlated to a blend morphology close to its final state (nodular-like with micronic dimensions), i.e. when the initial nanometric lamellar morphology is sufficiently altered due to dewetting. 

\subsection*{Effects of large amplitude oscillations}

\begin{figure}
\centering
\includegraphics[width=1.0\linewidth]{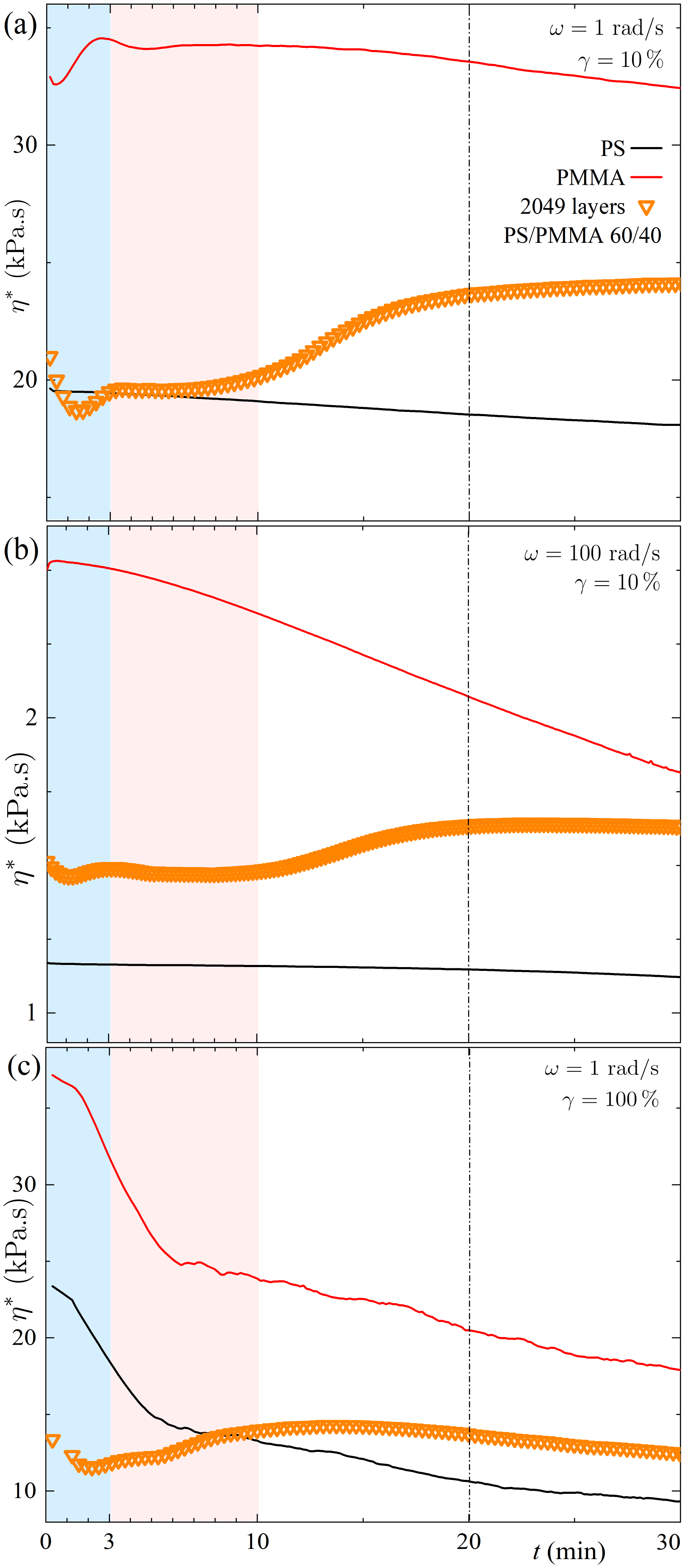}  
\caption{Complex viscosity, $\eta^*$, as a function of time at 180 \textdegree C for PS, PMMA and film with 2049 layers and composition 60/40 for various large strain amplitudes, $\gamma$, and  pulsations, $\omega$; (a) $\gamma=10$\%, $\omega=1$ rad/s, (b) $\gamma=10$\%, $\omega=100$ rad/s and (c) $\gamma=100$\%, $\omega=1$ rad/s.} 
\label{fig8}
\end{figure}

To explore further the effect of shear on dewetting or morphology change, additional LAOS measurements were performed on the 2049 layers 60/40 films. 
Tests with strain $\gamma=10$\% at $\omega=1$ and 100 rad/s were performed, as well as additional tests with $\gamma=100$\% at $\omega=1$ rad/s.
Figure \ref{fig8} shows $\eta^*$ as a function of time for these conditions. 

First, the rheological measurements with $\gamma=10$\% will be described. 
When $\omega=1$ rad/s, the viscosities of the neat polymers decrease slightly over the experiment time, with the average $\eta_{\text{PMMA}}^*$ equal to 33.7$\pm$0.7 kPa$\cdot$s, and 19$\pm$0.5 kPa$\cdot$s for PS. 
At $\omega=100$ rad/s, $\eta_{\text{PMMA}}^*$ decreases rapidly and linearly with time, while $\eta_{\text{PS}}^*$ remains stable. 
Surprisingly, the behavior of the multilayer film seems unaffected by this non-linearity and again, three time zones may be distinguished, analogous to the measurements with low strain $\gamma=0.1$\% in linear viscoelastic regime (LVER). 

When $\gamma=100$\%, see Fig. \ref{fig8}(c), the viscosities of the neat polymers decreases sharply and follow a power law behaviour with exponents around \mbox{-0.2} for $\eta_{\text{PMMA}}^*$ and \mbox{-0.24} for $\eta_{\text{PS}}^*$, respectively. 
The rheological behavior of the multilayer film also appears to change and increases to a ``second plateau'' observed before $t=10$ min. 
Then, the viscosity decreases slightly, similar to the neat components of the film. 


\begin{figure*}
\centering
\includegraphics[width=1.0\linewidth]{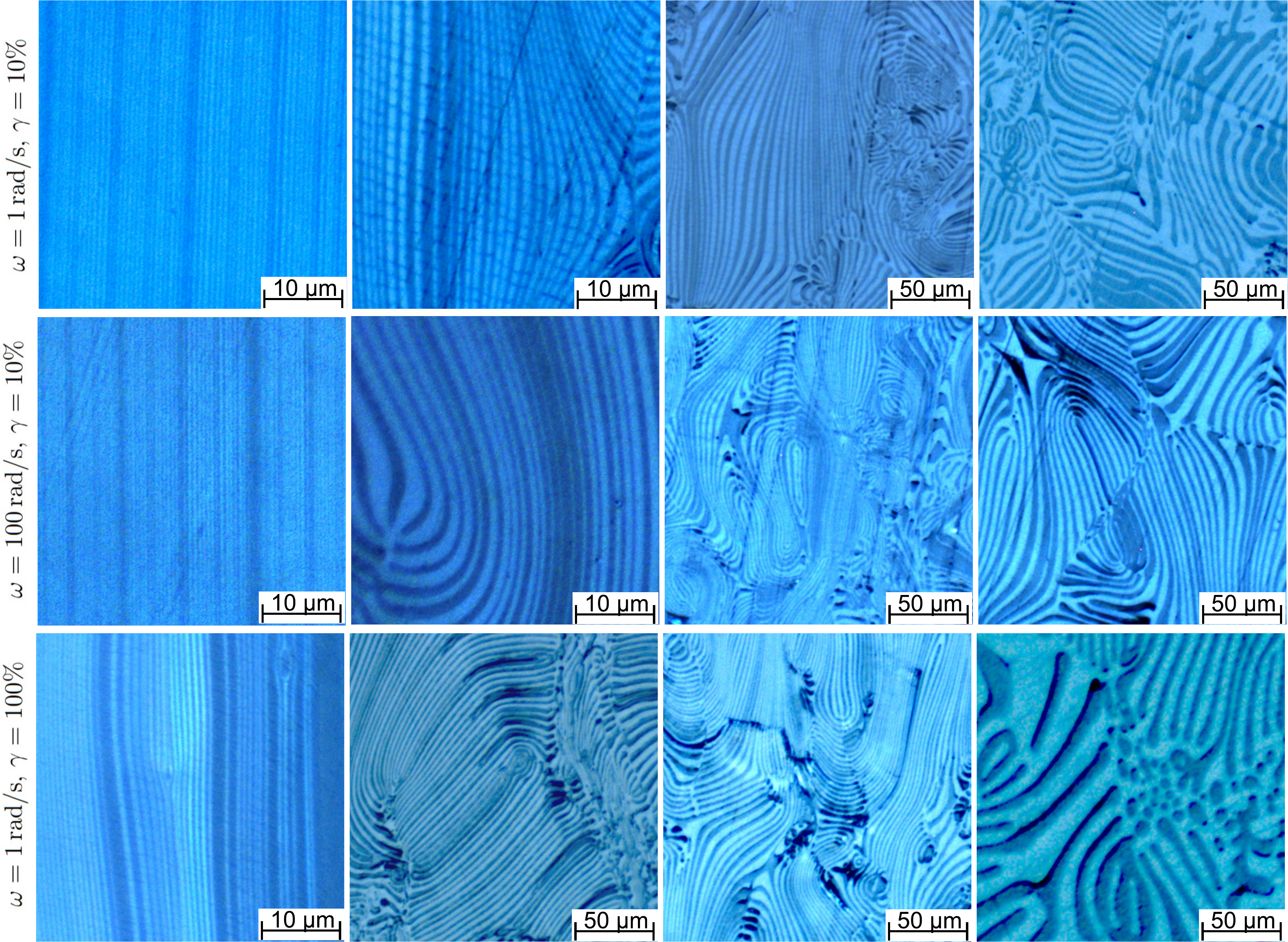}
\caption{The morphology of the 2049 layers 60/40 film after quenching for the different $\gamma$ and $\omega$ conditions. From left to right, $t=3$, 10, 20 and 30 min. Light blue represents PS and dark blue PMMA.}
\label{fig9}
\end{figure*}

Interrupted tests were then performed until $t=3$, 10, 20 and 30 min to observe the morphology of the samples in the time zones of interest, shown after quenching in Fig. \ref{fig9}. 
Both the films tested with $\omega=1$ and 100 rad/s with $\gamma=10\%$ appear completely stable after $t=3$ min, i.e., no layer breakup was observed. 
After $t=10$ min, layer breakups were observed at both frequencies, similar to what was observed with $\gamma=0.1\%$. 
However, there is a clear difference between the samples measured with $\omega=1$ and 100 rad/s within the LVER, and the samples presented in Fig. \ref{fig9}, measured outside of the LVER.
Clearly, the LAOS leads to large deformations of the layers. Moreover, the thickness of the layers is bigger than the ones of the samples tested with $\gamma=0.1\%$. 
The multilayer structure is still present, though distorted due to the instabilities causing the neighboring layers to move. 
After $t=20$ min, the morphology of the sample is even more disorganized and the multilayer morphology is mostly lost. 
As previously, this transient morphology is followed by the creation of a lamellar-like blend, with few droplets present, after $t=30$ min for both samples.

\begin{figure}
\centering
\includegraphics[width=1.0\linewidth]{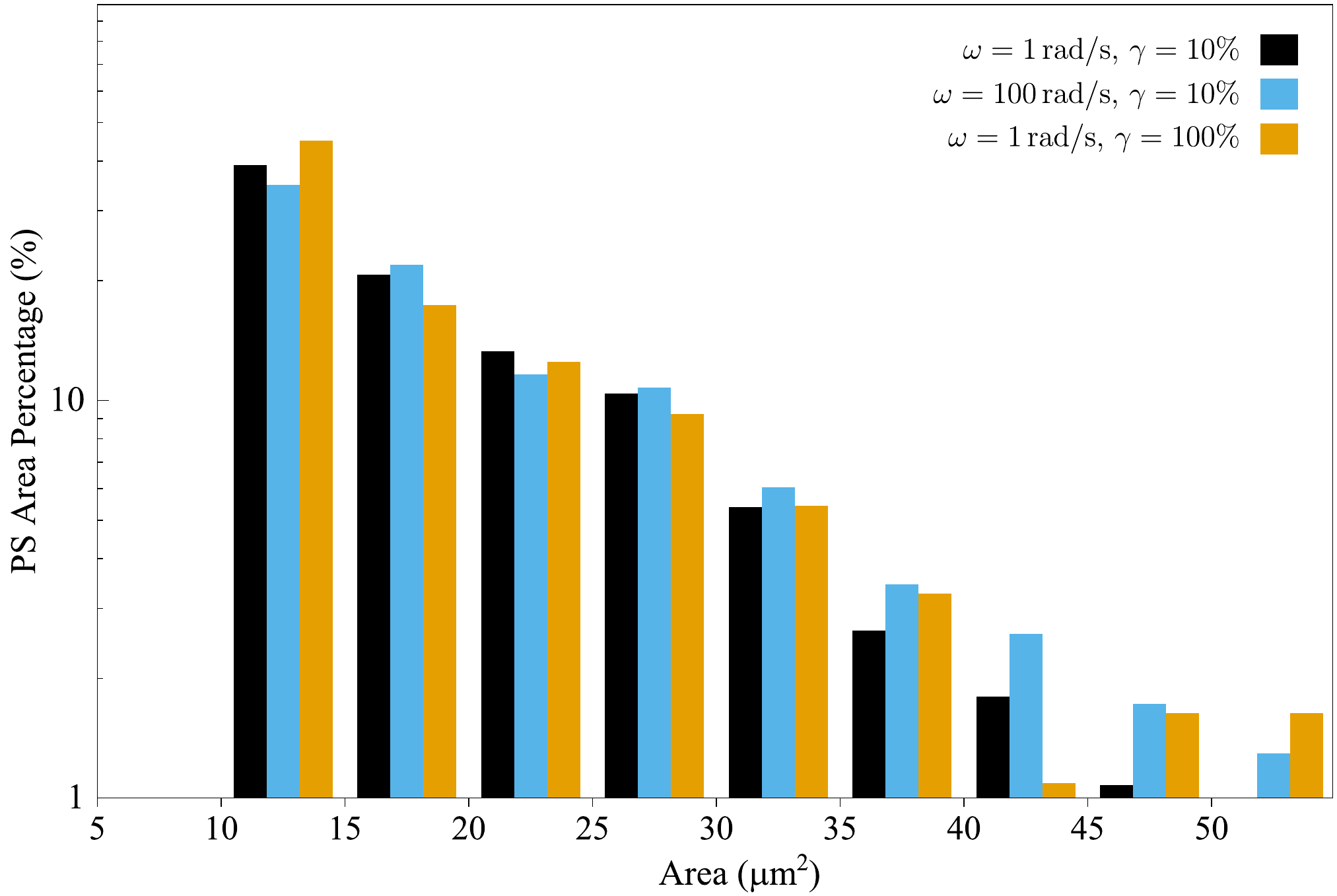}
\caption{Area distribution of drops  with circularity
$c>0.5$ in the cross-section of films with 2049 layers, composition 60/40 PS/PMMA, after $t=30$ min measurements in the non-linear regime. 
The bin size of each sample is 5 $\mu$m$^2$.} 
\label{fig10}
\end{figure}

Figure \ref{fig10} presents the PS area distribution with circularity $c>0.5$ of the samples at $t=30$ min for the different LAOS conditions. 
The area distribution of all drops is also similar, as presented in Fig. S6 in SI.
Again, a dominating droplet area around 15 $\mu$m$^2$ is obtained independently of the LAOS conditions tested. 

However, what is interesting is the comparison of the percentage of the drops with $c>0.5$, this time not with $\omega$, but with $\gamma$, as shown in Tab. \ref{tab1} (last 3 lines to compare with the second line as a reference). 
With an increasing strain, the percentage of PS nodules decreases, meaning that large strains stabilize the lamellar-like morphology, at least in the measured experimental time window. This can be explained by the fact that large external deformations applied during dewetting will prevent droplet formation and favor elongated structures, though shorter and thicker than the initial multilayer morphology.

\section*{Conclusions}

Multi-nanolayer coextrusion is an intriguing process in a sense that it allows, in one step, the fabrication of films made of thousands of continuous and alternating layers of immiscible polymers, all having nanometric thicknesses. Considering the high temperature and relatively large residence time involved during extrusion, at such thicknesses the layered morphology should be unstable because of dewetting phenomenon. As recalled in the introduction, it was suggested \citep{Bironeau2017},  based on 2D-simulations \citep{Davis2010,Kadri2021}, that shear occurring during extrusion could be a stabilizing factor helping to obtain continuous layers. However, a recent 3D-simulation study \citep{Dhaliwal2024} showed that dewetting could still occur perpendicularly to the applied shear, resulting in no significant increase of the time before layer breakup occurrence. 
The initial objective of this study was then to develop a non-direct method to capture the onset of dewetting in multilayer films, to gain further insight on these questions from an experimental perspective. 

As a validation of the previous hypotheses, for thick multilayer films after, the shear SAOS complex viscosity remains constant, well described by a simple additivity rule and is associated to unbroken layers.
For submicrometric layers, a drop in complex viscosity has been evidenced at short times.
In  all tested oscillatory conditions, and though it cannot be quantitatively related to a rupture time that would be the signature of layer breakup, it is clearly linked with changes in the films' morphology at the early dewetting stages. 
As expected, from the organized multilayer structure, some layers start to spontaneously break up. This causes the neighboring layers to move, causing even higher distortions in the still-organized structure. From that, the broken layers retract into thicker and shorter lamellae, along with a coalescence phenomenon, and a lamellar-like blend is observed. The final dewetting step is the break up of these thick layers into drops, reminiscent of the Plateau-Rayleigh instability \citep{Papageorgiou1995}. After $t=30$ min, a similar blend morphology composed of lamellar and nodular structures has indeed been observed at all tested $\omega$ and $\gamma$. 
The second plateau in the rheological signal, occurring from about 20 min, can then be correlated with the global loss of the organized multilayer structure, with a gradual change from the multi-nanolayer structure to the lamellar-like, followed by a nodular-like blend with dispersed phases of micronic sizes.

However, the fractions of the lamellar-like and drop morphologies differ from one experimental condition to another. When the pulsation is the same ($\omega=1$ rad/s) but the strain increases, the amount of drops at a given time (30 minutes here) decreases. At small strains, the amount of drops depend on the initial composition and on the pulsation: if no particular trend can be observed as a function of pulsation for the 60/40 composition where both PS and PMMA layer thicknesses are initially on the order of 200 nm, it is clearly observed that higher pulsations favor the nodular morphology for the 30/70 composition where PS layers are initially much smaller than the PMMA ones (about 100 and 400 nm respectively). This seems in agreement with previous experimental work focusing this time not on the onset of nucleation but on the growth of holes in ultra-thin polymer layers, which showed that higher shear rates led to higher dewetting speed \citep{Dmochowska2022}.  

As a future work, a quantitative statistical study using AFM rather than optical microscopy (hence with a higher resolution) can help providing a more precise link between morphology changes and rheological behavior, and the possibility to use macroscopic tools such as rheometer to characterize quantitatively nanoscale phenomena such as dewetting of ultra-thin polymer layers. The role of phase elasticity (through the second normal stress difference) on the morphological stability shall also be studied in more details, as suggested in previous works \citep{Vanoene1972,VanOene1978}.

\subsection*{Supporting Information}
The online version contains supplementary
material available at https://doi.org/10.1007/sxxxxx-xxx-xxxxx-x.

\subsection*{Acknowledgements}
The Ecole Doctorale SMI (ED 432) is acknowledged for granting A.D. the fellowship for her Ph.D. 
We acknowledge Dr. Alain Guinault for his help during the multilayer coextrusion.
\subsection*{Data Availability}
The data sets used and analysed in this study are available from the corresponding author on reasonable request.
\section*{Declarations}
\section*{Competing Interest}
The authors declare that they have no known competing financial interests or personal relationships that could have appeared  to influence the work reported in this paper.
\section*{Conflicts of Interest}
The authors declare that they have no conflicts of interest.
\bibliography{references}
\end{document}


\title[Article Title]{Supplementary Information on

Transient rheology and morphology in sheared nanolayer polymer films}

\author{\fnm{Anna} \sur{Dmochowska}}
\author*{\fnm{Jorge} \sur{Peixinho}}
\author{\fnm{Cyrille} \sur{Sollogoub}}
\author*{\fnm{Guillaume} \sur{Miquelard-Garnier}}
\affil{\orgdiv{Laboratoire PIMM}, \orgname{CNRS, Arts et Métiers Institute of Technology, Cnam}, \city{Paris}, \country{France}}
\maketitle

\renewcommand{\thefigure}{S1}
\begin{figure}
\centering
\includegraphics[width=0.84\linewidth]{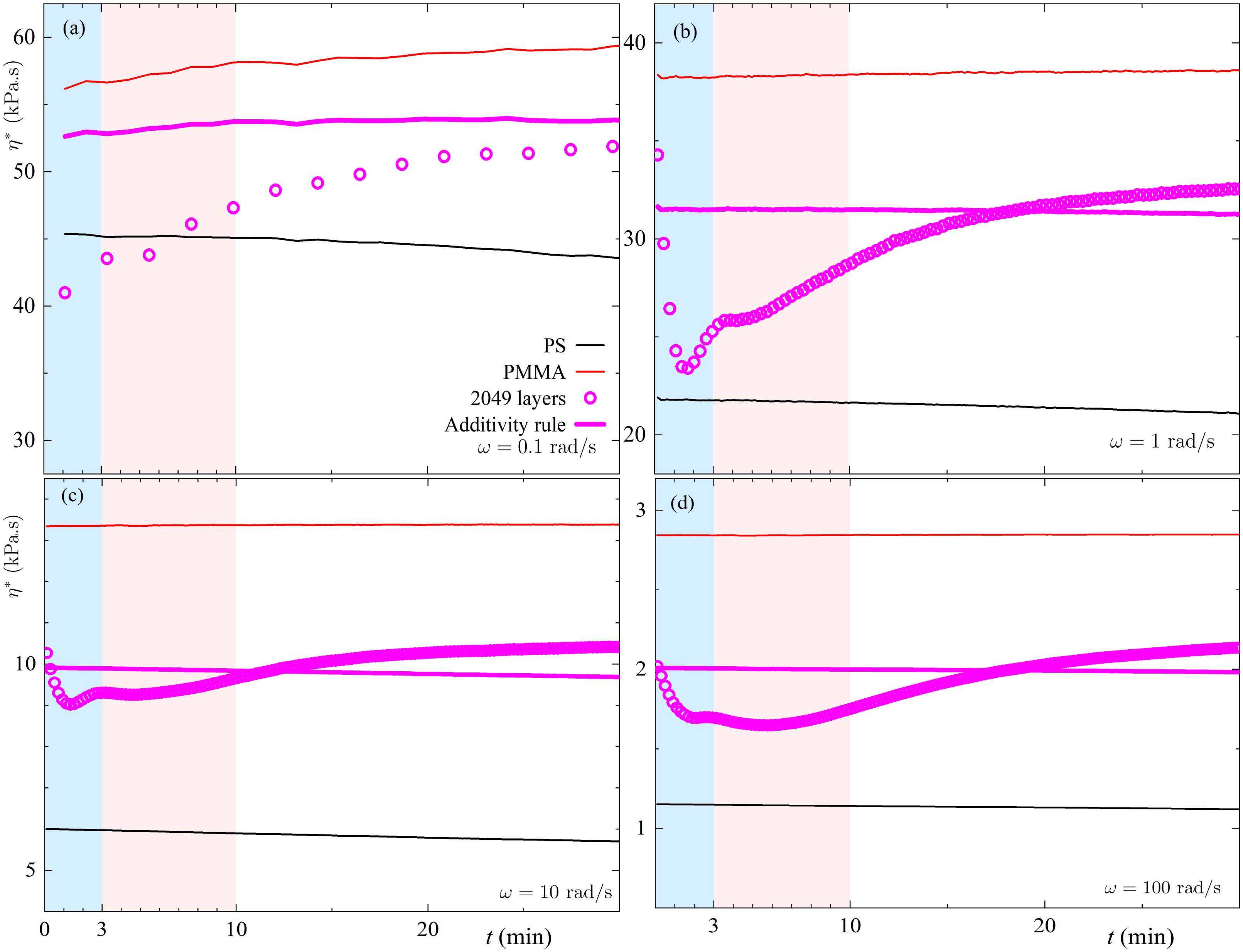}
\caption{The complex viscosity, $\eta^*$, as a function of time at 180 \textdegree C for PS, PMMA and 2049 layers film PS/PMMA 30/70. Shear strain $\gamma=0.1$\% for (a) $\omega=0.1$, (b) 1, (c) 10 and (d) 100 rad/s. The pink symbols are the measured data while the thick line is $\eta^*$ calculated from the additivity rule.}
\end{figure}

\renewcommand{\thefigure}{S2}
\begin{figure}
\centering
\includegraphics[width=1.0\linewidth]{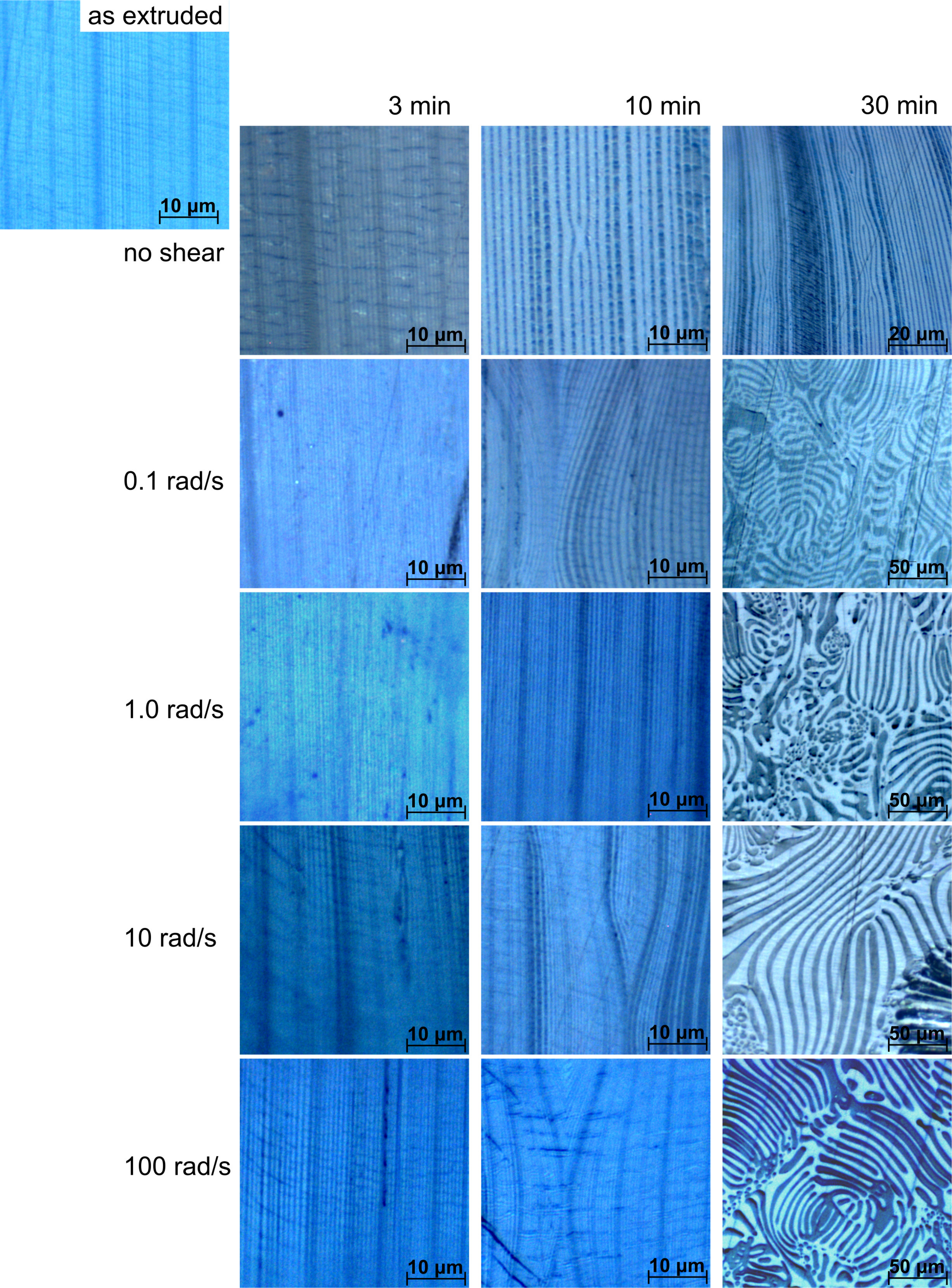}
\caption{Optical microscope images of 2049 layers 60/40 PS/PMMA multilayer film's cross-sections. 
The image on the top left corner is the extruded morphology.
First row, static conditions. 
Following rows, cross-sections after $\gamma=0.1\%$, and $\omega=0.1$, 1, 10, and 100 rad/s (from top to bottom) at 3, 10 and 30 min (from left to right)}  
\end{figure}

\renewcommand{\thefigure}{S3}
\begin{figure}
\centering
\includegraphics[width=1.0\linewidth]{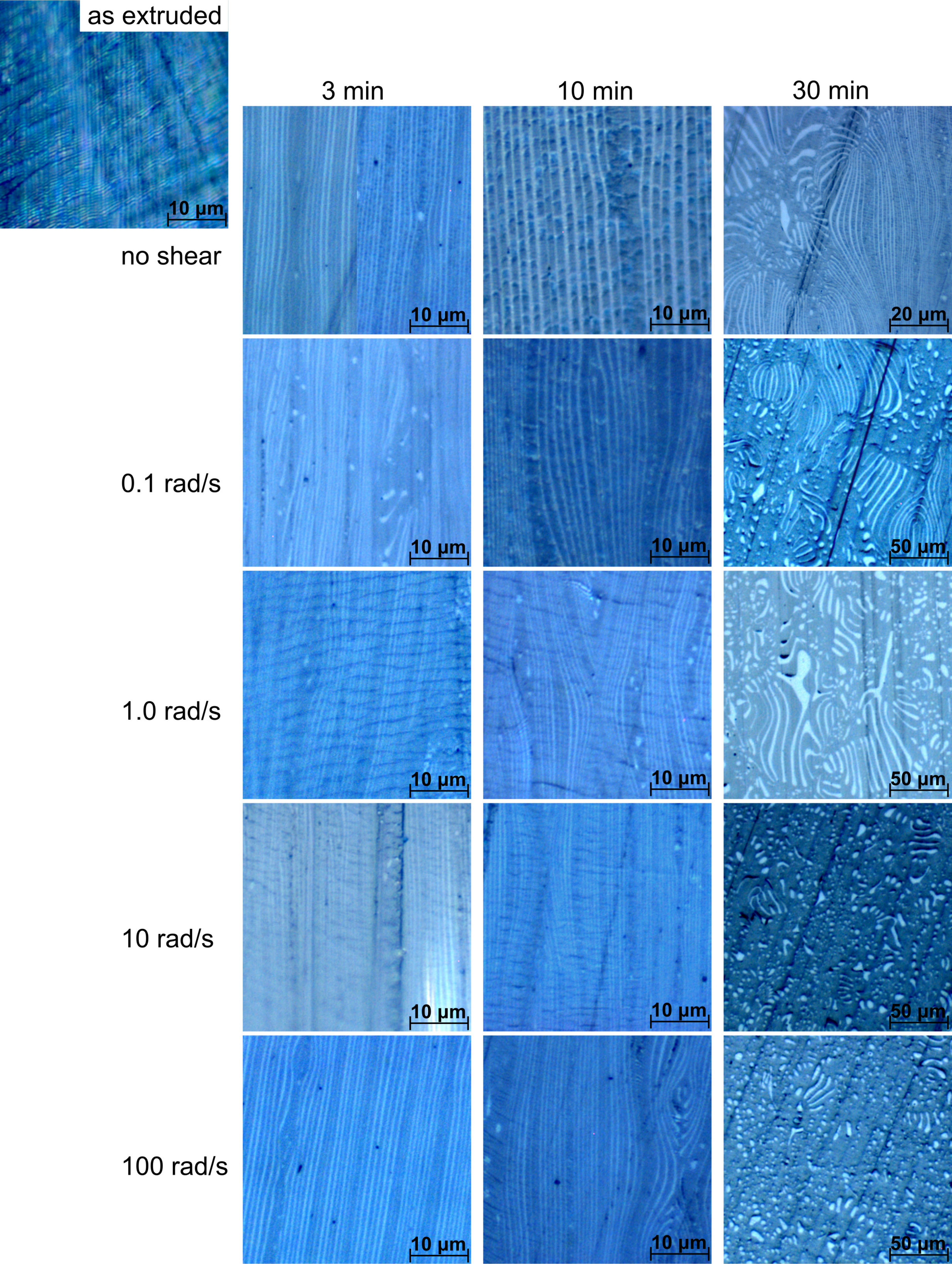}
\caption{Optical microscope images of 2049 layers 30/70 PS/PMMA multilayer film's cross-sections. 
The image on the top left corner is the as extruded morphology.
First row shows the morphology evolution with time without shear applied. 
Following rows, cross-sections after $\gamma=0.1\%$, and $\omega=0.1$, 1, 10, and 100 rad/s (from top to bottom) at 3, 10 and 30 min (from left to right)} 
\end{figure}

\renewcommand{\thefigure}{S4}
\begin{figure}
\centering
\includegraphics[width=1.0\linewidth]{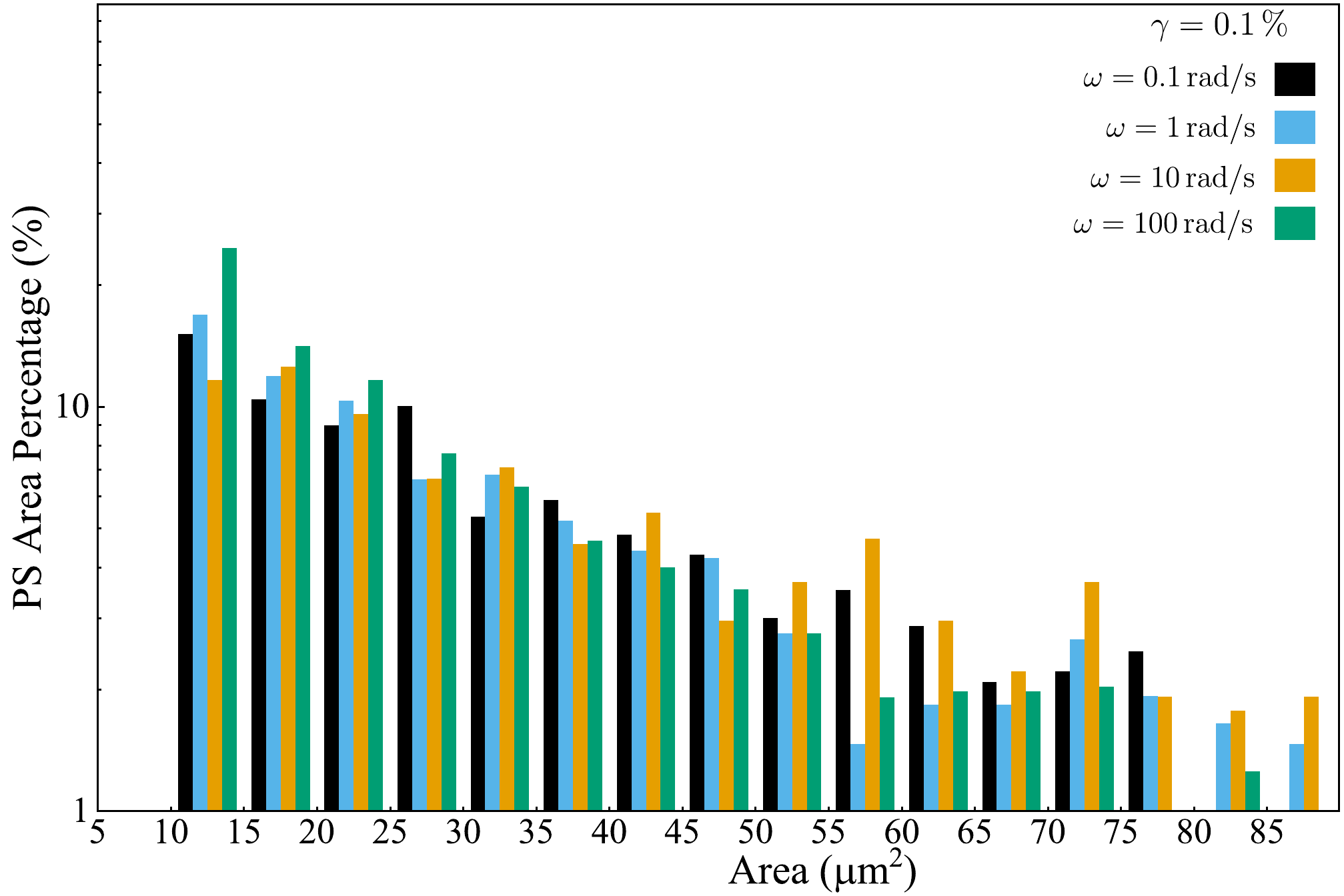}
\caption{Area distribution of all PS drops (with any circularity) in the cross-section of films with 2049 layers, composition 60/40 PS/PMMA, after $t=30$ min for $\gamma=0.1\%$ and $\omega=0.1$, 1, 10 and 100 rad/s. The bin size of each sample is 5 $\mu$m$^2$.} 
\end{figure}

\renewcommand{\thefigure}{S5}
\begin{figure}
\centering
\includegraphics[width=1.0\linewidth]{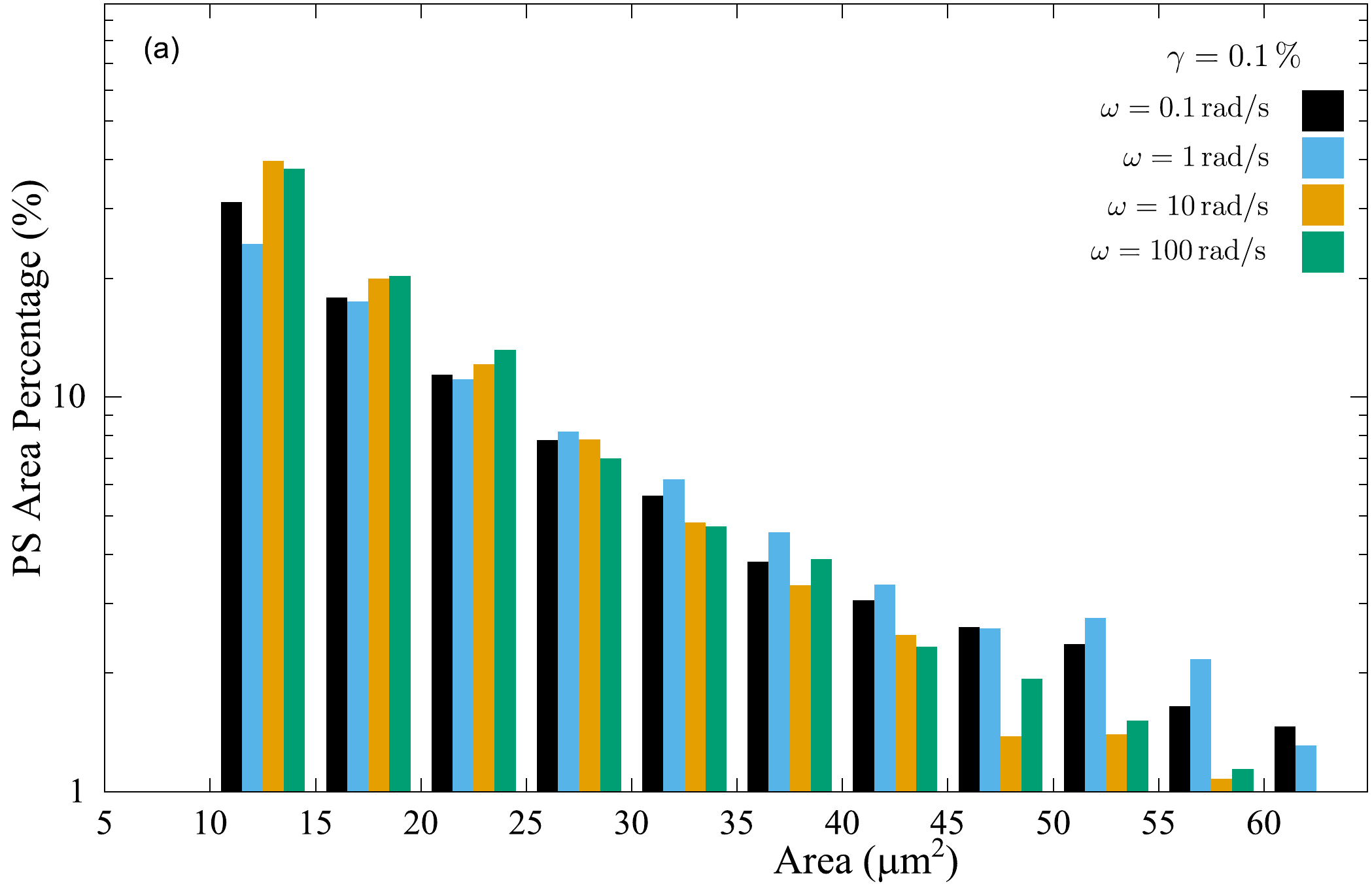}
\includegraphics[width=1.0\linewidth]{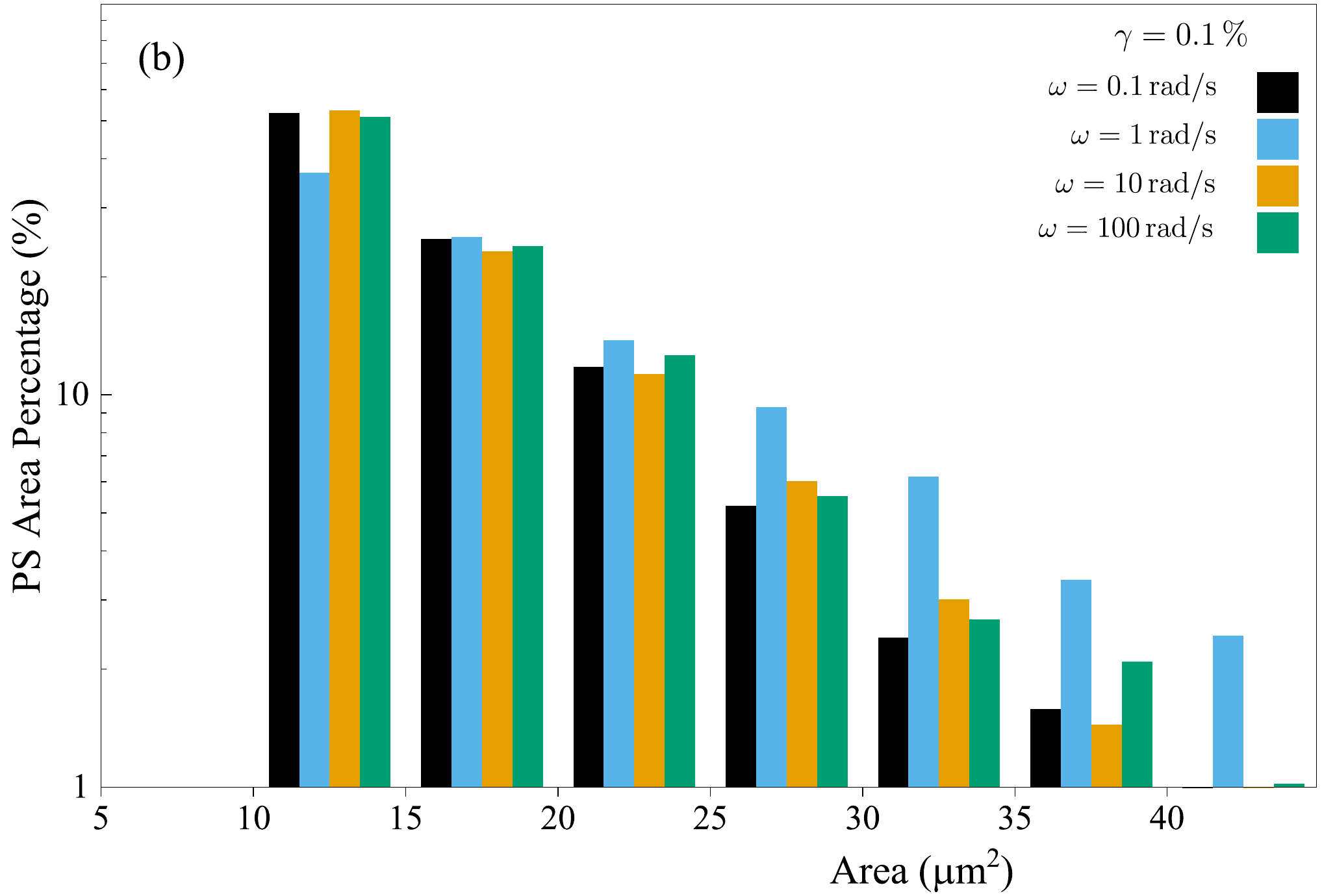}
\caption{Area distribution of (a) all PS drops (with any circularity) and (b) PS drops of circularity $c>0.5$ in the cross-section of films with 2049 layers, composition 30/70 PS/PMMA, after $t=30$ min for $\gamma=0.1\%$ and $\omega=0.1$, 1, 10 and 100 rad/s. The bin size of each sample is 5 $\mu$m$^2$.} 
\end{figure}

\renewcommand{\thefigure}{S6}
\begin{figure}
\centering
\includegraphics[width=1.0\linewidth]{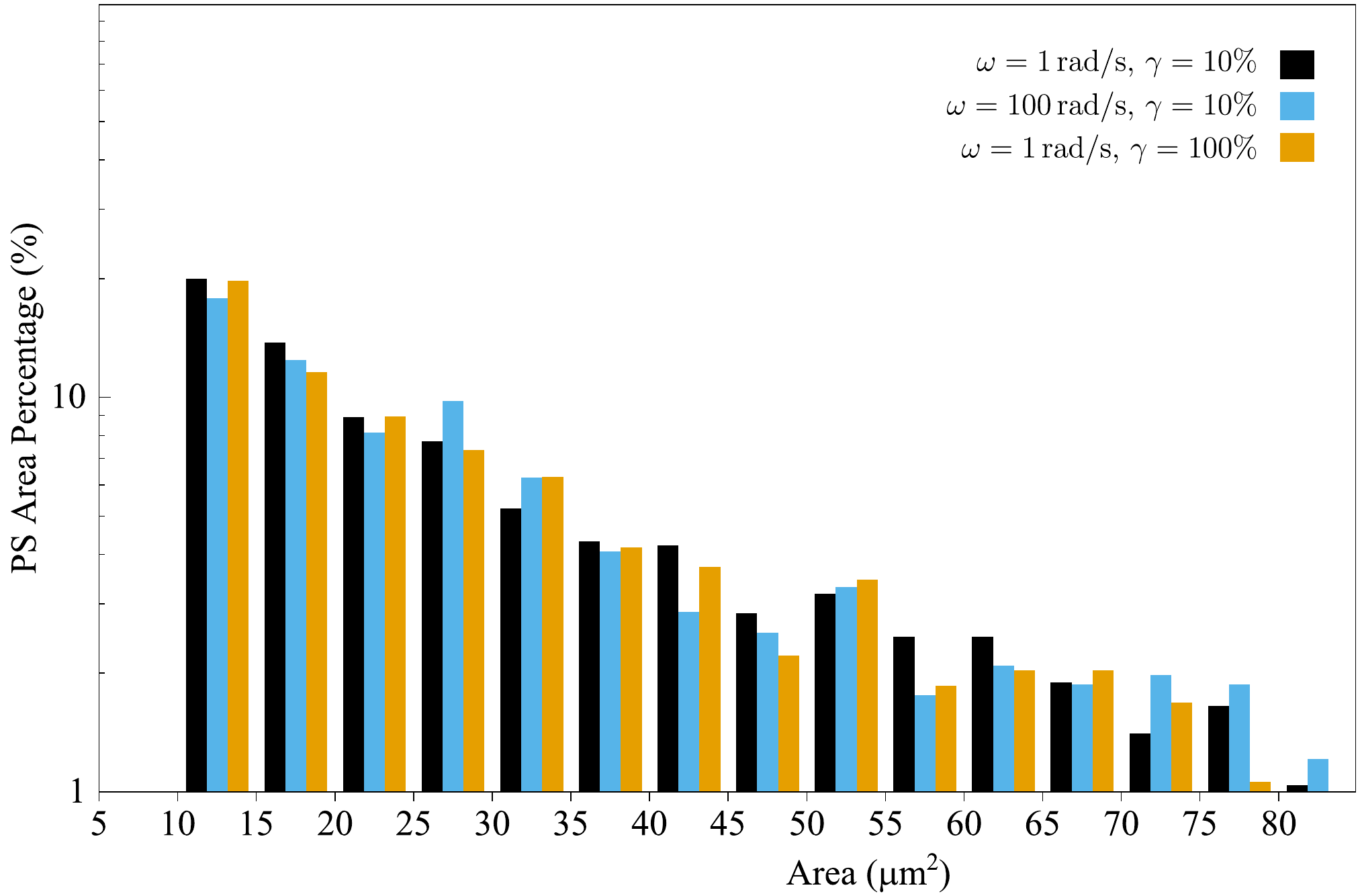}
\caption{Area distribution of all PS drops (with any circularity) in the cross-section of films with 2049 layers, composition 60/40 PS/PMMA, after $t=30$ min for $\gamma=10$ and 100\%. The bin size of each sample is 5 $\mu$m$^2$.} 
\end{figure}